# Nanoconfinement Effects on Intermolecular Forces Observed via Dewetting

**Authors**

Tara (Tera) Huang [1], Evon S. Petek[1], and Reika Katsumata[1]*

[1]Department of Polymer Science and Engineering, University of Massachusetts Amherst

120 Governors Dr, Amherst, MA, 01003, USA

*Corresponding author: Reika Katsumata

**Email:** rkatsumata@umass.edu

**Author Contributions:** R.K., T.H., and E.P. contributed to the conceptualization of the study. T.H. was responsible for data collection and analysis. R.K. and T.H. drafted and edited the manuscript, with E.P. contributing to the editing. All authors participated in manuscript review. R.K. supervised the project and secured funding. All authors contributed to the development of methodology.

**Abstract**

Although wettability is a macroscopic manifestation of molecular-level forces, such as van der Waals (vdW) forces, the impact of nanoconfinement on material properties in reduced film thickness remains unexplored in predicting film stability. In this work, we investigate how nanoconfinement influences intermolecular interactions using a model trilayer system composed of a thick polystyrene (PS) base, a poly(methyl methacrylate) (PMMA) middle layer with tunable thickness (15-95 nm), and a 10 nm top PS film. We find that the dewetting behavior of the top PS layer is highly sensitive to middle PMMA thickness, deviating from classical vdW-based predictions that assume bulk material properties. By incorporating nanoconfinement-induced changes in PMMA refractive index into the calculation of the Hamaker constant, we present a modified theoretical framework that successfully captures the observed behavior. This study links dewetting behavior and material property change as a function of underlayer thickness, providing direct evidence that nanoconfinement in soft matter systems significantly influences long-range intermolecular interactions. We show that film stability can be tuned solely by adjusting underlying layer thickness, while preserving both chemistry and thickness of top functional film. This finding carries broad implications for thin-film technologies across scientific and engineering disciplines by enabling performance-targeted interface design.

## Introduction

With technological advancements, the precise coating of polymer films less than 100 nm in thickness is necessary for device miniaturization and enhanced optical transparency across various industries[1–3], including packaging[4], sensors[5], and electronics[6]. However, as thickness decreases, films tend to be unstable, posing challenges to maintaining conformal coatings over an extended period[1,7,8]. Such instabilities result in a phenomenon known as dewetting, where a uniform polymer film spontaneously retracts from its substrate and ruptures into droplets, stripes, or pillars[7–11]. Although dewetting is often regarded as a source of defects in applications that require uniform coatings[12], conversely, the dewetting instabilities can also be leveraged to create patterned materials[13], potentially avoiding the high cost of lithography[14], facilitating the fabrication of micro-scaffolds[15], or reducing raindrop adhesion on windshields[16]. Acknowledging both the challenges and opportunities, prominent technological relevance has cultivated various efforts to suppress or enhance dewetting[13], including end-group functionalization[17], substrate modification[18], and nanoparticle[19].

In contrast to approaches that require altering the chemical composition of the substrate and films, previous work has focused on controlling the stability of polymer thin films via intermolecular interactions. The stability of a dewetting film is governed by the interfacial free energy of the system, $F(h)$, where variations in the film and underlayer coating thickness can significantly impact wetting and dewetting outcomes[24,25]. The interfacial energy is determined by both short-range interactions, such as hydrogen bonding, and long-range interactions, which are due to van der Waals (vdW) forces. When the film thickness $h$ is sufficiently large, long-range interactions dominate, and the short-range interactions can be neglected. For a multi-layer film, like the example shown in **Fig. 1(a)**, the long-range interaction is defined as

$$F_{vdw}(h) = [(A_{33} - A_{23})/6\pi h^3] - [(A_{13} - A_{23})/6\pi(h+d)^3] \quad (1)$$

where each $A$ is the Hamaker constant defining vdW interactions between the film with thickness $h$ (Layer 3) and the substrate (Layer 1) in a medium with distance $d$ (Layer 2)[23,26]. The expression of $F_{vdw}(h)$ indicates that dewetting outcomes can vary significantly even for films of the same polymer material and thickness due to underlayer effects from Layer 1 and 2. Utilizing this vdW potential, Seemann et al. demonstrated that the interfacial potential can effectively explain the stability of thin polystyrene (PS) films (Layer 3) on top of Si wafers (Layer 1) with varying silicon oxide layer thickness $d$ (Layer 2)[11]. A subsequent study by Sun et al. demonstrated that the characteristic wavelength of the spatial correlation ($\lambda$), which represents the wavelength of the most unstable and fastest-growing patterns, can be adjusted by merely changing the thickness of the underlayer polymer coating ($d$) also showing an underlayer effect[22]. In addition to film stability, underlying layers have also been reported to show a 'remote' influence on other physical properties, such as glass transition temperature ($T_g$)[27–33] and crystallization behavior[34]. Those observations underscore the broad and significant impact of underlayer effects on various polymer properties.

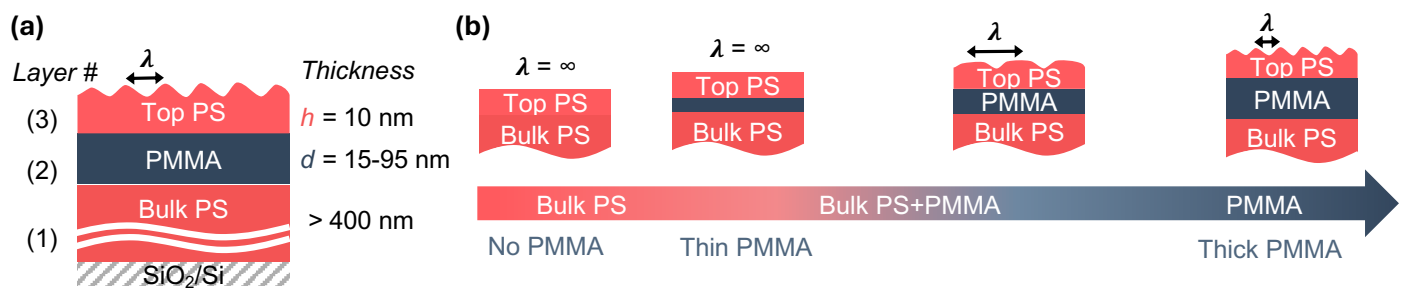

**Fig. 1 (a)** Illustration of the tri-layered system used in this study. (1) Bottom layer: Bulk PS (> 400 nm thick). (2) Middle layer: PMMA ($d$ = 15 - 95 nm). (3) Top layer: Dewetting PS ($h$ = 10 nm). **(b)** Our hypothesis regarding the vdW and nanoconfinement effect of the PMMA underlayer on system stability ($\lambda$) as its thickness ($d$) ranges from 0 to infinity.

While previous research has significantly advanced our understanding of how underlayer effects control dewetting behavior, the impact of nanoconfinement on this remains underexplored. Nanoconfinement in polymer films refers to the phenomenon where the material properties begin to deviate from their bulk values as the film thickness decreases, often around 100 nm thick, depending on the experimental output[35–37], including $T_g$[27–33], dielectric constant[37], refractive index[38], and many others[39]. Given the extensive research on nanoconfinement in polymers, it is expected that the changes in physicochemical properties would, in turn, affect intermolecular interactions. For example, Li *et al.* reported that the water contact angle on PS films started to decrease at thickness around ~ 120 nm, and the authors rationalized this thickness dependence due to the change of PS refractive index as a function of film thickness[35]. Our previous study further investigated the impact of polymeric underlayer thickness on surface wettability, using contact angle measurements on bilayer system of cross-linked PS and poly(methyl methacrylate) (PMMA) with top layer thickness varied from 5 to 100 nm for two case studies: PMMA(top)/PS(bottom) and PS(top)/PMMA(bottom)[40]. When the PMMA layer thickness on top of PS is around 10-20 nm, the contact angle of vdW-dominated ionic liquid becomes nearly identical to that of the PS underlayer, as if PMMA layer is invisible in terms of wettability, termed wetting transparency similar to reported work on graphene[41]. In contrast to PMMA, the PS layer on a bulk PMMA substrate did not exhibit wetting transparency, likely due to the dominance of nanoconfinement effects on PS surface wetting. However, to our knowledge, such nanoconfinement effects have not been considered in the context of the intermolecular force landscape in multi-layer polymer films.

In this study, we hypothesize that in the nanoconfinement regime, long-range intermolecular interactions (vdW) and effects from the underlayer will differ from those observed in bulk materials. We aim to test this hypothesis by developing a multi-layered model system using PS (hydrophobic) and PMMA (hydrophilic), which are commercially available and well-studied, as illustrated in **Fig. 1(a)**. If the hypothesis is true, one would expect that the influence of the PMMA underlayer diminishes once its thickness is thin enough, resulting in system stability comparable to that of a thin layer of PS on bulk PS (*i.e.*, stable with no dewetting), as shown in **Fig. 1(b)**. The bulk PS layer (> 400 nm thick) is designed to block the influence of the silicon substrate and systematically study the interaction only between polymer films. The high molar mass of the PMMA layer prevents undesired interdiffusion and mixing between dewetting PS and PMMA layers during the annealing process at 140 °C (at least 35 °C higher than bulk $T_g$) for a maximum of 72 hours under a high vacuum (< $10^{-6}$ Torr)[22]. To minimize the potential influences of non-equilibrium state of thin films due to residual stress generation during spin coating[42], a constant spinning speed was used across all PMMA film thicknesses in our study. In other words, variations in PMMA thickness were achieved by adjusting the solution concentration rather than the rotation speed. This multi-layered system design effectively isolates the specific effects of polymer-polymer interactions and nanoconfinement on film stability, characterized by a *time-independent* characteristic wavelength ($\lambda_0$) of the dewetting patterns. The $\lambda_0$ values are then used to calculate the system's stability as the PMMA layer thickness ($d$) varies. Notably, we find that the nanoconfinement-induced changes in the refractive index of PMMA explain our experimental observations using the recalculated effective Hamaker constants. The recalculated effective Hamaker constants take into account variations in the PMMA underlayer film thickness and nanoconfinement effects, thereby providing a more accurate model for predicting stability and dewetting behavior. Our results highlight the critical role of both the underlayer and nanoconfinement effects in influencing intermolecular interactions and determining the stability of multi-layered thin polymer films.

**Results and Discussion**

The multi-layered model system was fabricated as follows, detailed in **Methods** and **Fig. S1**. A bulk layer of PS (bulk PS layer, $M_w$ = 35 kDa, $Đ$ = 1.72) is first spin-coated onto the pre-treated and cleaned silicon substrate at a thickness greater than 400 nm. A high molar mass PMMA (PMMA layer, $M_w$ = 828 kDa, $Đ$ = 1.09) is prepared on top of the bulk PS layer using water floating from mica, for thicknesses varying from 15 nm to 95 nm. Lastly, a 10 nm thick PS layer (dewetting PS layer, $M_w$ = 35 kDa, $Đ$ = 1.72) is water-floated on top of the PMMA layer.

A series of control experiments verified that the PMMA middle layer and bottom PS bulk layer (> 400 nm thick) remained stable and uniform throughout the annealing process at 140 °C under vacuum, and that the top 10 nm PS also remained stable in the absence of the PMMA layer. As shown in **Fig. 2**, a control bilayer sample (PMMA control) is designed with the thinnest layer ($d$ = 15 nm), representing the most unstable PMMA thickness in our experiment. This sample exhibited stable (*i.e.*, no dewetting) behavior throughout the annealing period, confirmed by optical microscope (OM) in **Fig. 2** and atomic force microscopy (AFM) with the root-mean-square film roughness = 0.57 nm in 20×20 μm² scanning areas (**Fig. S2**). Similarly, a second control bilayer sample (PS control in **Fig. 2**) shows that the top PS layer does not undergo dewetting upon annealing in the absence of a middle PMMA layer. These control experiments indicate that the observed changes in the dewetting behavior of the top PS layer are governed by the middle PMMA layer thickness in our system.

In **Fig. 2**, the surface morphology of 10 nm-thick PS layers with different PMMA underlayer thickness $d$ was carefully studied as a function of annealing time at a constant temperature 140°C. The OM images reveal distinct morphological changes over time. To obtain quantitative analysis, the $2\pi/\lambda$ of dewetting PS was determined by Fast Fourier Transform (FFT) of OM images in **Fig. 2** using the *ImageJ* program as detailed in **Methods** and **Supplementary Information**. Rather than relying on visual inspection, the identification of early and late stages (enclosed by the red dotted line in **Fig. 2**) is based on a change in the slope of the reciprocal characteristic wavelength ($2\pi/\lambda$) versus time plot, as discussed later and **Fig. 3(a)**.

While AFM offers higher resolution, we opted for OM, following the established image analysis process to evaluate system energy, as described by Seemann *et al.*[11] and Sun *et al.*[22], both of whom utilized OM images due to their ability to capture a larger observable area. To validate the reliability of using OM for dewetting morphology analysis, we cross-checked the OM and AFM results on an AFM control sample, confirming that both methods yield identical FFT peak positions (**Fig. S5**).

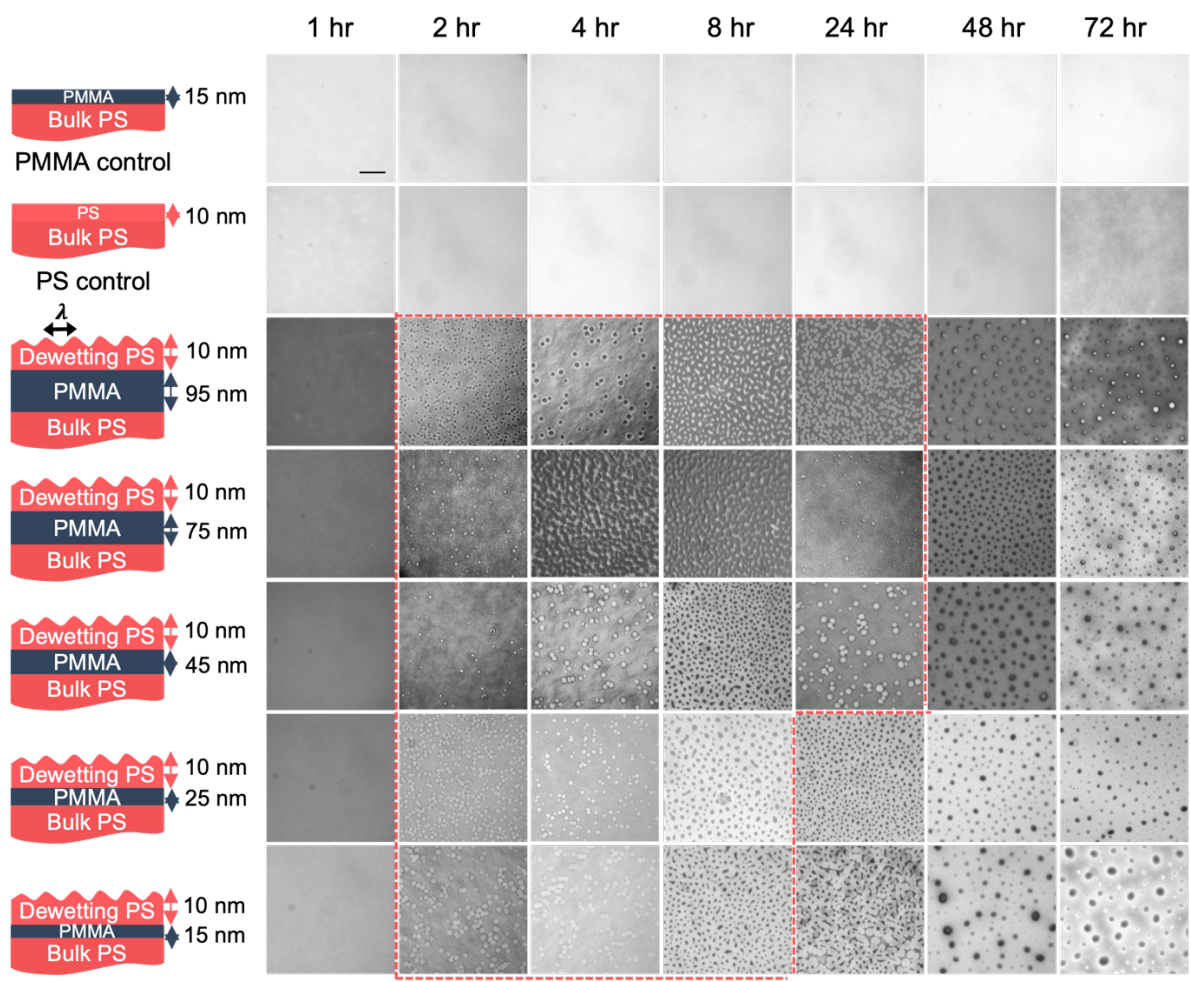

**Fig. 2** Optical microscopy images showing distinct morphology at different time points at 140°C. The images within the red dotted line represent the early stage of dewetting, as identified through the quantitative analysis shown in **Fig. 3(a)** and **Fig. S6**, rather than by visual inspection alone. The bilayer sample (PMMA control) ensured the stability of both the bulk PS and the middle PMMA layers throughout the entire annealing process. The second bilayer sample (PS control) confirms that the top PS layer remains stable without PMMA underlayer. The scale bar represents 10 μm and the height and width of all images are 50.71 μm.

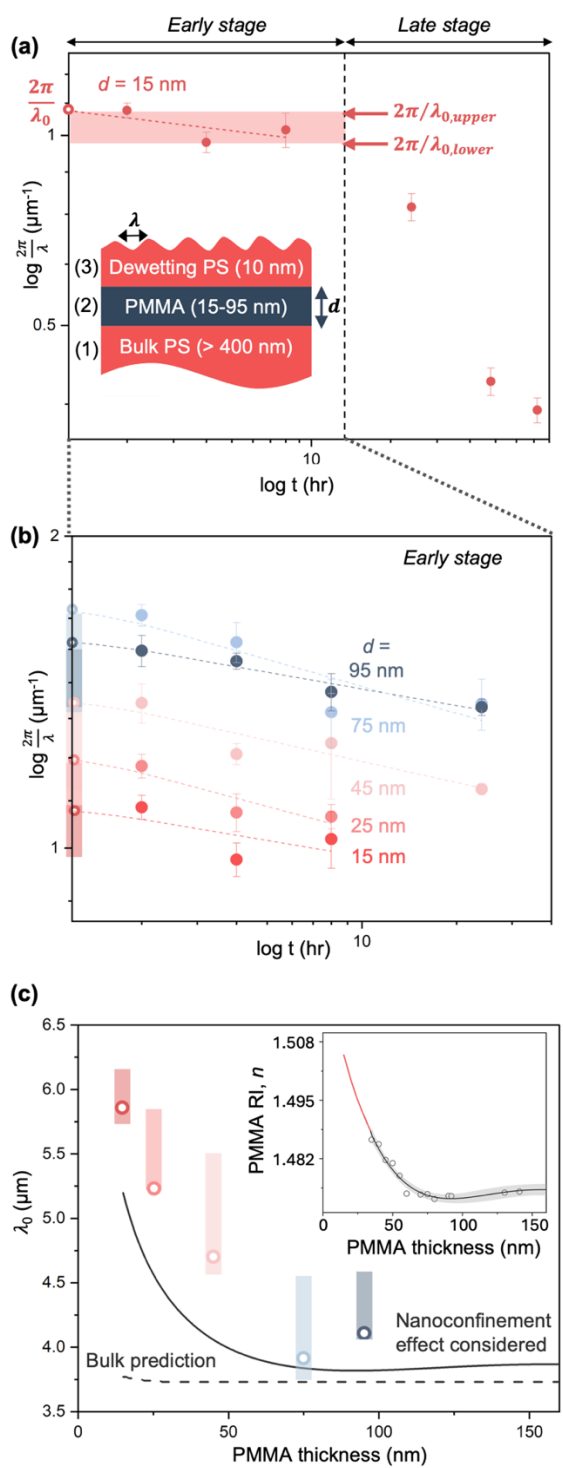

**Fig. 3 (a)** Log-log scale of $2\pi/\lambda$ as a function of annealing time ($t$) at 140 °C with PMMA thickness $d$ = 15 nm. The dashed line represents the early stage of dewetting, where the time-independent characteristic wavelength, $\lambda_0$, is defined by the y-intercept of the linear regression. The upper and lower bounds of $2\pi/\lambda_0$ are defined from the minimum and maximum values of $2\pi/\lambda$ observed during the early stage. The error bars for each time point are from standard deviation of four measurements from one individual sample. **(b)** The determination of $2\pi/\lambda_0$ for $d$ = 15-95 nm. **(c)** Dependence of $\lambda_0$ on PMMA thickness ($d$). The experimental data points of $\lambda_0$ for each thickness are represented by open circles, with upper and lower bounds indicated by shaded rectangles, as determined from **Fig. 3(b)**. Theoretical calculations using bulk properties (dashed line) cannot predict the change in system instability with varying PMMA thickness. In comparison, this change is captured better when the nanoconfinement effects with changing refractive index (RI) in the PMMA layer are considered using the effective Hamaker constant (solid line). The inserted figure shows the relationship between RI and $d$ adapted from Han et. al[38].

Two stages of dewetting are further confirmed from the $2\pi/\lambda$ versus annealing time plot in **Fig. 3(a)** and **Fig. S6**, consistent with previous literature[21,43]. In the early stage of the dewetting process (highlighted by the red dotted line in **Fig. 2)**, the $\lambda$ remains nearly constant over the annealing time interval, while the late stage of dewetting occurs when the holes grow large enough and begin to coalesce, causing $\lambda$ to increase dramatically over the annealing time. We rigorously measure the $\lambda$ at various annealing time intervals for up to 72 hours, avoiding the arbitrary selection of $\lambda$ at a single time point. **Figure 3(a)** shows that the two-stage dewetting process is evident from a change in the slope of the relationship between $\log(2\pi/\lambda)$ and time, taking PMMA thickness $d = 15$ nm as an example. The time-independent characteristic wavelength, $\lambda_0$, is determined from the y-intercept of the linear regression of all the $(2\pi/\lambda)$ in the early stage (red dashed line), where the upper bound and lower bond (depicted as a shaded rectangle) of $2\pi/\lambda_0$ are the maximum and minimum $\lambda$ values observed during the early stage of dewetting. This data analysis was applied to other samples with $d = $ 15-95 nm, as shown in **Fig. 3(b)** and late-stage data for all other $d$ is available in **Fig. S6**. In this study, we focus exclusively on the early stage of dewetting with time-independent characteristic wavelength to exclude the influences of dewetting kinetics, which is also reported to have underlayer-thickness dependences[44,45], and distinguishing between spinodal dewetting and nucleation and growth solely through OM images remains a topic of debate[46].

The plot of $\lambda_0$ with varying PMMA layer thicknesses $d$ is depicted in **Fig. 3(c)**, where the upper and lower bounds were determined by the maximum and minimum $\lambda$ values observed during the early stage of dewetting in **Fig. 3(b)**. The $\lambda_0$ values increase as PMMA thickness decreases, suggesting that the system becomes more stable with thinner PMMA layers (as $\lambda_0$ approaches infinity for a uniform and non-dewetting film). This change is consistent with our hypothesis in **Fig. 1(b)** and can be explained by the underlayer effect from bulk PS, where long-range interactions penetrate the PMMA layer and stabilize the dewetting PS. However, we hypothesize that the PMMA-thickness-dependent instability of top PS is governed not only by conventional vdW force predictions but also by nanoconfinement effects within the PMMA layer.

To assess the presence of nanoconfinement effects in the PMMA layer within our system, we first conduct conventional predictions for $\lambda_0$ representative of bulk behavior. The silicon substrate's vdW influence can be safely disregarded because vdW interactions from the silicon substrate are significant only for PS film thicknesses below 20 nm[31], which is much smaller than the thickness of PS bottom layer (400 nm) in this study. Consequently, we derive the effective Hamaker constant based on a tri-layer system consisting of bulk PS, PMMA, and the top PS layer, following previous literatures[23,47]:

$$\lambda_0 = 4\pi(\pi\gamma_3)^{1/2}h^2[(A_{33} - A_{23}) + (1 + d/h)^{-4}(A_{23} - A_{13})]^{-1/2} \qquad (2)$$

where 1, 2, and 3 refer to the layer numbers shown in **Fig. 1(a)**, and $\gamma_3$ is the surface tension of the top PS film (30.8 mN/m)[11]. Assuming the intermolecular forces (represented by the Hamaker constants in this equation) remain the same as their bulk values, we first calculate the Hamaker constant for pure PS ($A_{11}$, $A_{33}$) and PMMA ($A_{22}$) substance using Lifshitz's equation[25] (**Eq. 3**):

$$A_{11} = \frac{3}{4}kT\left(\frac{\epsilon_1-1}{\epsilon_1+1}\right)^2 + \frac{3h\nu_e}{16\sqrt{2}}\frac{(n_1^2-1)^2}{(n_1^2+1)^{3/2}} \qquad (3)$$

where $k$ is the Boltzmann constant (1.38 × 10$^{-23}$ J K$^{-1}$), $T$ is room temperature (298 K), $h$ is the Plank constant (6.63 × 10$^{-34}$ J Hz$^{-1}$), and $\nu_e$ is the main electronic adsorption frequency constant of a material (typically 3 × 10$^{15}$ Hz). The refractive indices of bulk PS ($n_1$), PMMA ($n_2$), and top PS ($n_3$) were literature values measured from bulk materials, where $n_1 = n_3 = 1.572$ and $n_2 = 1.475$[38]. The dielectric constants ($\varepsilon$) of PS and PMMA are calculated by $\varepsilon = $

$n^2$. The Hamaker constant for bulk PS/top PS ($A_{13}$) and PMMA/top PS ($A_{23}$) in air can be further calculated by pairwise approximation $A_{12} = \left(\sqrt{A_{11}} - \sqrt{A_{air}}\right)\left(\sqrt{A_{22}} - \sqrt{A_{air}}\right)$, where the Hamaker constant for air ($A_{air}$) is $0^{25}$. The dashed line in **Fig. 3(c)** shows the predicted $\lambda_0$ based on the Hamaker constants for bulk materials in **Eq. 2**. While bulk prediction matches well with the experimental $\lambda_0$ values for the large $d$, the bulk model underpredicts $\lambda_0$ when the underlying PMMA thickness is reduced.

Nonetheless, the bulk prediction provides reliable predictions for $\lambda_0$ when the substrate is inorganic. When **Eq. 2** is applied to the well-established polymer-on-inorganic substrate system[11]: Si/SiO$_x$/PS and data with bulk values ($A_{\text{Dewetting PS/Dewetting PS}} = A_{33} = 7.09 \times 10^{-20}$ J, $A_{\text{Si/Dewetting PS}} = A_{13} = -1.3 \times 10^{-19}$ J, $A_{\text{SiOx/Dewetting PS}} = A_{23} = 2.20 \times 10^{-20}$ J; details in **Supplementary Information** and **Fig. S7**) shows the bulk prediction based on **Eq. 2** can explain the instability of PS on SiO$_x$ much better than PS on PMMA. This assessment suggests that for soft material substrates, using bulk behaviors to predict film stability cannot fully capture the intermolecular interactions in the thin film regime. In other words, nanoconfinement effects within the PMMA layer may play a crucial role in determining interactions between multilayer bodies, especially in polymer thin films where geometrical nanoconfinements dominate film properties.

Notably, when the nanoconfinement effect is considered, the trend for the predicted $\lambda_0$ (**Fig. 3(c)**, solid line) aligns more closely with experimental data than the bulk prediction. The bulk predictions of the $\lambda_0$ value in **Eq. 2** were refined by incorporating the effective Hamaker constant ($A_{23}$), as detailed in **Supplementary Information**. The effective Hamaker constant captures nanoconfinement by incorporating changes in the PMMA refractive index due to the confined environment, thereby reflecting the modified vdW interactions at nanoscale distances. The effective Hamaker constant for pure PMMA substance ($A_{22}$) is now recalculated as a function of $d$ using Lifshitz's equation in **Eq. 3**[25]. The refractive indices of PMMA ($n_2$) as a function of thickness ($d$) were adopted from Han *et al.*'s work[38], shown as an inserted figure in **Fig. 3(c)**, yielding the effective Hamaker constant for PMMA/Dewetting PS in air ($A_{23}$) from pairwise approximation in the same way as the bulk prediction.

Among several potential factors that could improve the accuracy of the nanoconfinement-based predictions, we identify three key considerations: (1) interfacial effects between PS and PMMA[48], (2) different trends in PMMA refractive index ($n_2$)-thickness ($d$) relationships[38,49,50], and (3) nanoconfinement effects on PS. First, the interfacial width between PMMA and a 10 nm PS film is reported to be approximately 1.5 nm[48], accounting for only 15% of the thickness of the top PS layer in our system. Given this relatively small proportion and the experimental challenges in directly probing PS/PMMA interfacial effects[48], we assume that the polymer chain exhibits potential anisotropy at the PS interface. Although this assumption may not be comprehensive, it provides a practical foundation for our current predictive model and can be refined as more detailed interfacial information becomes available in the future. Second, we recognize that the thickness dependence of the refractive index of PMMA ($n_2$) has been a subject of ongoing debate[38,49,50], which shows different trends shown in **Fig. S8**. This study adopted Han *et al.*'s data[38] because of the careful treatment of potential experimental artifacts in ultrathin polymer film measurements by ellipsometry without inherent assumptions, such as uniformity and homogeneity of the polymer layer and a constant Hamaker constant. Although Han *et al.* reported PMMA refractive index values that deviate significantly from the commonly cited 1.495, this discrepancy is consistent with observations from El-Nasser *et al.*, who found that PMMA's refractive index depends strongly on molecular weight[51]. Lastly, the current instability prediction could be further improved by incorporating the nanoconfinement effects within the PS dewetting layer itself, given its thickness is only

10 nm. However, due to spatial variations in film thickness caused by dewetting, accurately determining both the refractive index and local thickness remains experimentally challenging.

**Conclusions**

In conclusion, we developed a multi-layered polymer thin film system to investigate molecular interactions through dewetting. Stability predictions based solely on bulk material properties failed to fully capture the observed variations in soft matter systems, unlike those on hard substrates such as $SiO_x$. By accounting for the nanoconfinement effects of the refractive index on the effective Hamaker constant, we acknowledge that the dielectric properties directly influence how electromagnetic interactions, such as van der Waals forces, propagate through materials, and our new treatment of the dewetting model provides a more accurate representation of the observed experimental behavior. This finding highlights the importance of accounting for nanoconfinement effects in the study of thin film dewetting and offers a new strategy for stabilizing polymer multilayer systems by changing underlayer thickness without changing materials or chemistry.

**Materials and Methods**

*Solution preparation.* Polystyrene (PS, $M_w$ = 35 kDa, $M_w / M_n$ = 1.72, Aldrich, product number: 331651-25G) and poly(methyl methacrylate) (PMMA, $M_w$ = 828 kDa, $M_w / M_n$ = 1.09, Agilent, product number: PSS-MM850K) solutions of varying concentrations were dissolved in toluene (Thermo Fisher Scientific) to make thin films with the desired thickness. All the solutions were then placed on an orbital shaker at 120 RPM overnight to achieve equilibrium. To ensure purity, the solutions underwent double filtration using membrane filters (Fisherbrand PTFE Membrane Filter, 0.45 μm for PMMA and 0.22 μm for PS) before the spin coating process.

*Wafer Treatment before Spin Coating.* Silicon wafers (1 cm × 1 cm) (University Wafer, Silicon 100 mm P/B (100) 0−100 Ω-cm 500 μm SSP TEST, 3 × 3cm) underwent a thorough cleaning process starting with nitrogen gas to eliminate dust, and then washing by sonication (Fisherbrand CPX3800 Ultrasonic Bath) with 5% (v/v) CC-54 Detergent Concentrate (Thermo Fisher Scientific), DI water (two times), acetone (Thermo Fisher Scientific), and isopropyl alcohol (Thermo Fisher Scientific) each for 10 minutes. Subsequently, the wafers were air-dried in an oven at 120°C for 30 min and then subjected to a 15-minute oxygen-plasma treatment (Jelight Model 18 UVO Cleaner). The freshly cleaned wafers underwent an initial spin coating process (Laurell WS-650-23 B) with toluene at 6000 rpm for 30 seconds.

*Polymer Films Fabrication.* The first bulk PS layer (> 400 nm) was deposited through spin coating immediately following the initial toluene treatment with a 4 wt% solution at 1500 rpm for 60 seconds. The second PMMA and third PS layers were spin-coated onto mica sheets at 4500 rpm for 60 seconds (SPI-Chem Mica, Grade V-5) and subsequently transferred onto the prepared underlayer by floating on deionized water. Before transferring the third PS layer, the films were allowed to sit in ambient air to facilitate water removal. Following the PS layer transfer, the films were left in ambient air overnight and later subjected to a 24-hour period in a custom-made high-vacuum oven ($10^{-6}$ Torr) at room temperature to effectively eliminate residual solvent and water. Subsequently, the films were annealing under vacuum at 140°C for varying durations (2, 4, 8, 24, 48, and 72 hours).

*Ellipsometry.* The film thickness of each layer was measured by Gaertner Scientific LSE Stokes ellipsometer (λ = 632.8 nm, 70° incident angle) after drying.

*Atomic force microscopy (AFM) Imaging.* The bilayer control sample (PMMA control), consisting of bulk PS and a 15 nm PMMA layer, was examined using atomic force microscopy (Bruker MultiMode AFM) to ensure the stability of these two underlayers after annealing at 140 °C for 72 hours. The root-mean-square (RMS) roughness of the film was measured using AFM over a 20×20 μm² scanning area, confirming the uniformity of the underlayers after annealing with an RMS roughness of 0.57 nm. Gwyddion software was used to apply a second-degree polynomial fit. The AFM control sample, comprising bulk PMMA and a 10 nm top dewetting PS layer, was scanned to analyze the surface dewetting morphology to cross-validate the reliability of OM for image analysis.

***Optical Microscope (OM) Imaging.*** After the samples were annealed for the desired time intervals, optical microscopy (Zeiss Axioscope 5MAT with Axiocam 305 color) was performed on either trilayer or bilayer polymer samples using 100x lenses (**Fig. 2**). The bilayer samples were prepared for control experiments. The AFM control was designed to verify the use of OM as a reliable method for image analysis. The PMMA control was used to ensure the stability of the PMMA layers throughout the entire annealing process. In this sample, the PMMA layer was designed to be the thinnest (15 nm) in our system, as thinner layers are generally more prone to instability. The PS control was used to confirm the stability of the top PS layer in the absence of middle PMMA, ensuring that the top PS layer would not undergo dewetting upon annealing. All the trilayer images were analyzed using *ImageJ* software, following the steps outlined in **Fig. S3** to generate the Fast Fourier Transform (FFT) images shown in **Fig. S4**: (1) Convert all RGB images to grayscale (32-bit) and subtract background, (2) apply FFT, (3) perform radial average to obtain the intensity profile.

**Acknowledgments**

T.H. and R.K. gratefully acknowledge support from NSF CAREER Award (DMR 2046606) and 3M Non-Tenure Faculty Award for partial support for this project. T.H. thanks the PSE 50 Alumni Fellowship (2022). The authors thank James Merrill for his instruction in the water-floating process, Prof. Thomas McCarthy and Prof. James Watkins for providing the use of their ellipsometers, Prof. Thomas P Russell and Dr. James Pagaduan for meaningful discussion on the wafer cleaning process, Dr. Zhan Chen for assistance with AFM, and Tarik Karakaya for his help with the preliminary dewetting studies.

Supplementary Information for:

# Nanoconfinement Effects on Intermolecular Forces Observed via Dewetting

Tara (Tera) Huang [1], Evon Petek[1], and Reika Katsumata[1,*]

[1]Department of Polymer Science and Engineering, University of Massachusetts Amherst

120 Governors Dr, Amherst, MA, 01003, USA

*Corresponding E-mail: rkatsumata@umass.edu

*Table of contents*

## A. Multi-layered Film Fabrication

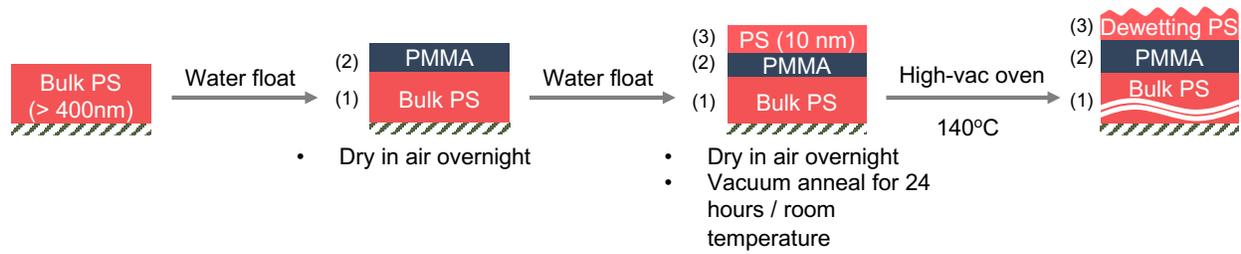

**Fig. S1** To achieve trilayer polymer films, a bulk Polystyrene (PS) layer is first spin-coated, followed by a water-floating technique for the middle Polymethylmethacrylate (PMMA) layer and the top PS layer. Each layer is thoroughly dried at room temperature before the next layer is deposited. After all layers are dried, the polymer films were vacuum annealed at room temperature for 24 hours. Following this, the temperature was increased to 140 °C to induce dewetting of the top PS layer.

## B. Ensuring Underlying Layers' Stability with Atomic Force Microscopy (AFM) Imaging

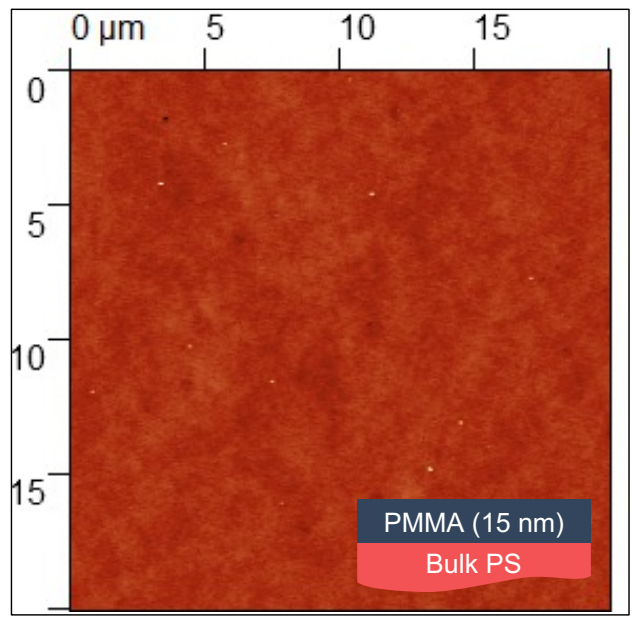

**Fig. S2** AFM image of the PMMA control sample after annealing at 140°C for 72 hours confirms the stability of both underlying layers throughout the entire annealing process (RMS = 0.57 nm).

## C. Time-points Study with Optical Microscope (OM) Imaging

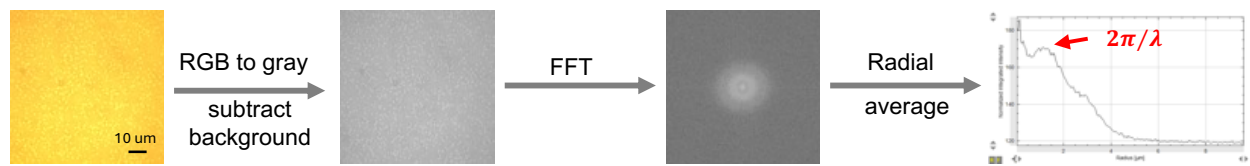

**Fig. S3** Procedure of Fast-Fourier Transformation (FFT) of the optical microscope images using the *ImageJ* program.

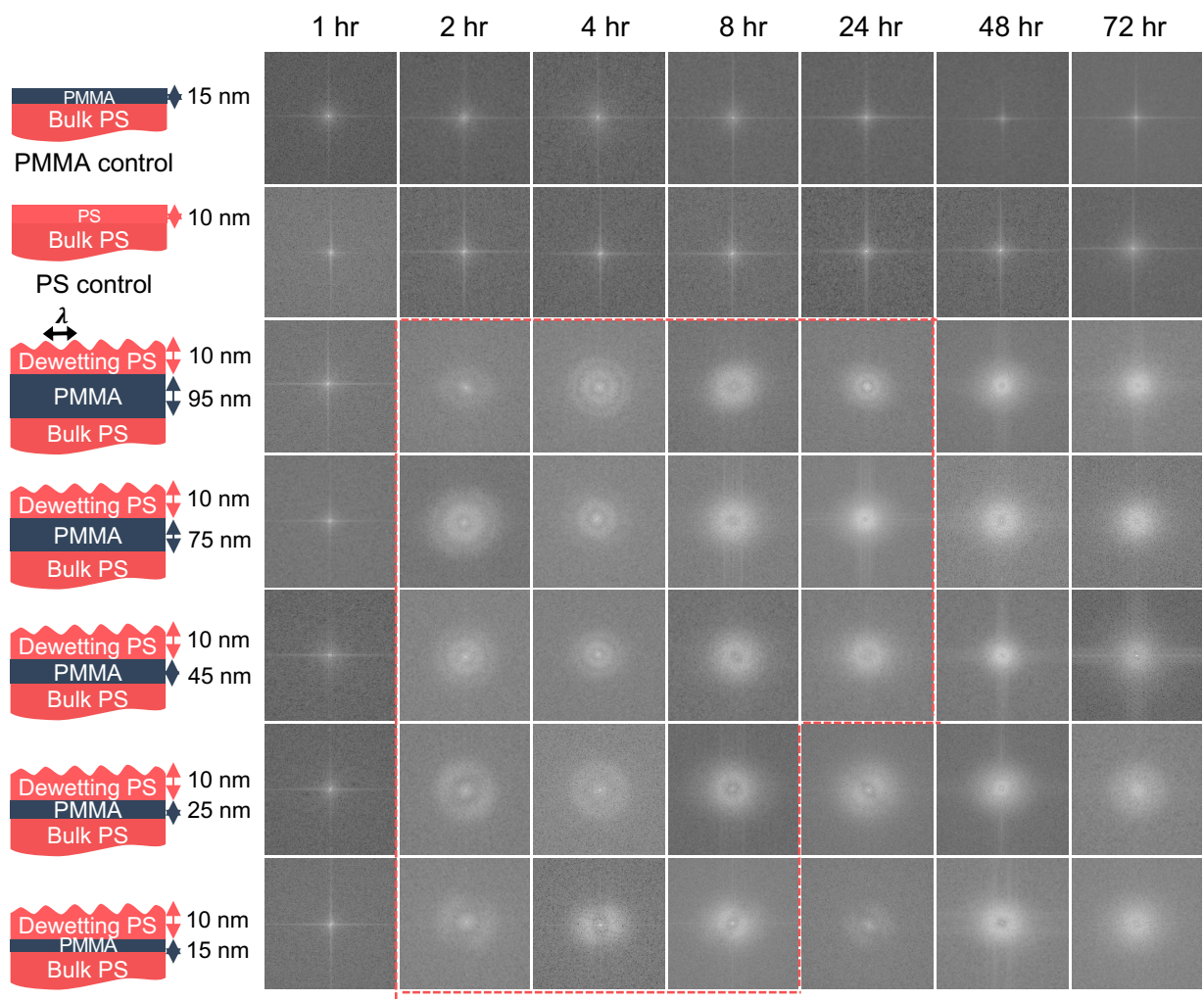

**Fig. S4** Fast-Fourier Transformation (FFT) of the OM images performed by *ImageJ* software. The images within the red dotted line represent the early stage of dewetting.

## D. Validation of OM Reliability via Cross-Comparison with AFM

For the AFM control bilayer sample, PS was dissolved in 1-chloropentane (Acros Organics), which acts as an orthogonal solvent to PMMA, allowing the PS layer to be directly spin-coated onto the PMMA layer without employing a water-floating transfer method.

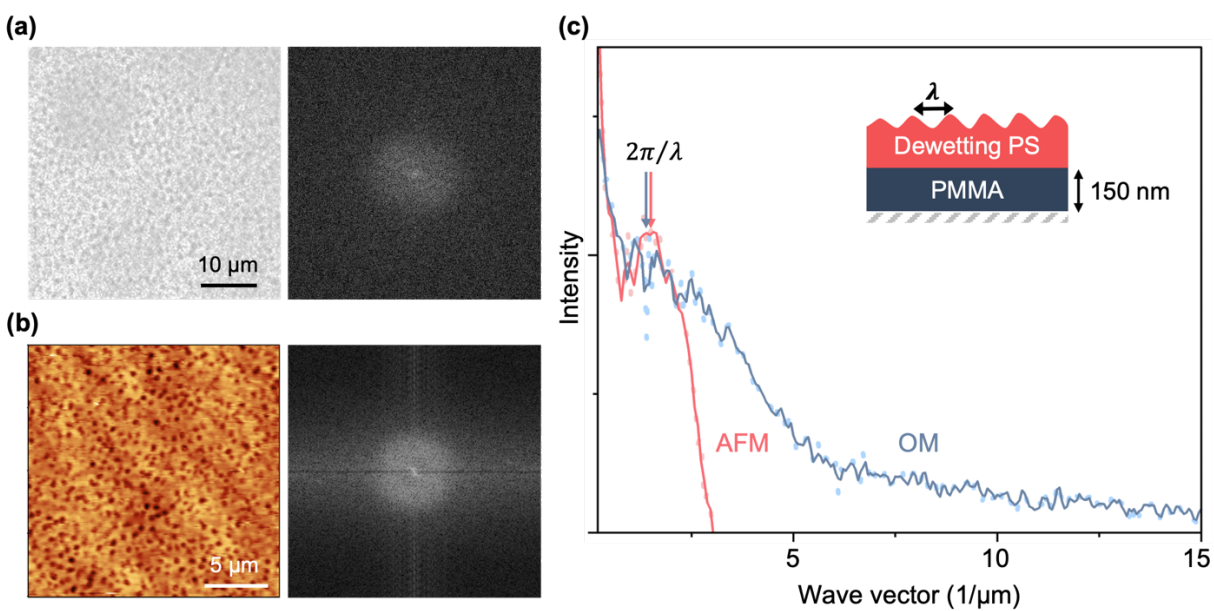

**Fig. S5 (a)** OM image of the AFM control sample and its corresponding FFT. **(b)** AFM image and its FFT. **(c)** Radial average profiles of the FFTs from OM (blue) and AFM (red) show identical peak positions, validating the reliability of OM for image analysis. Dotted lines represent raw data, while solid lines show the smoothed curves using the Savitzky–Golay method.

## E. Two-Stage Dewetting Behavior for Varying Polymethylmethacrylate (PMMA) Thicknesses

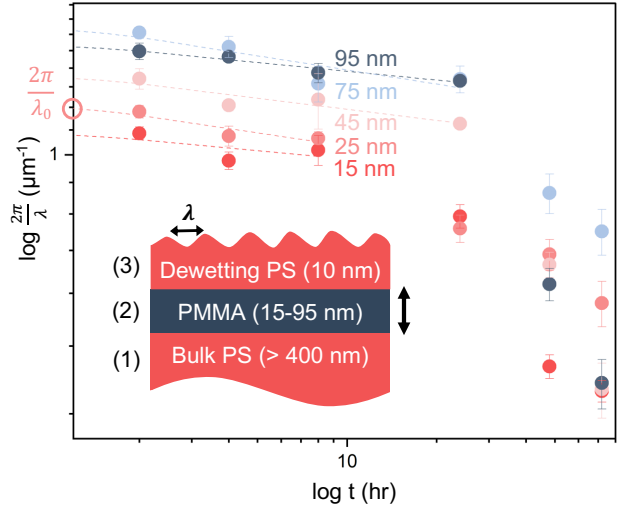

**Fig. S6** Log-log scale of $\frac{2\pi}{\lambda}$ as a function of annealing time at 140°C with varying PMMA thickness. The two-stage dewetting process is evident from the slope change in the linear fit. The dashed line represents the early stage of dewetting, where the time-independent characteristic wavelength, $\lambda_0$, is defined by the y-intercept of the linear regression. The error bars for each time point are defined by the standard deviation of four measurements from one individual sample.

## F. Analysis of Instability Prediction and Effective Hamaker Constant Calculation

### 1. Seemann's system

In **Fig. S7**, we confirm that **Eq. S1** effectively predicts instability using bulk properties when the underlayer is not a polymer with Seemann's system (Si/SiO/PS)[1]:

$$\lambda_0 = 4\pi(\pi\gamma_3)^{1/2}h^2[(A_{33} - A_{23}) + (1 + d/h)^{-4}(A_{23} - A_{13})]^{-1/2} \qquad (S1)$$

where $\gamma_3$ is the surface tension of the film, and 1, 2, and 3 refer to the layers from bottom to top in the different systems[2,3]. The bulk Hamaker constants were sourced from literature, where $A_{33} = 7.09 \times 10^{-20}$ (J), $A_{13} = -1.3 \times 10^{-19}$ (J), and $A_{23} = 2.20 \times 10^{-20}$ (J)[1,4].

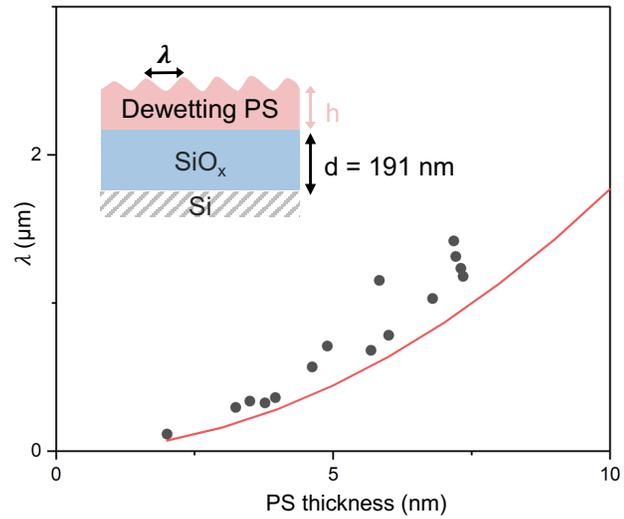

**Fig. S7** Validation of **Eq. S1** in predicting instability in Seemann's system. The experimental data (closed circles) shows strong agreement with the bulk prediction (red solid line).

2. **Multilayer polymer system**

The refractive index data (RI, open circles) was sourced from Han *et al.*'s work, and data points for thicknesses less than 150 nm were fitted with a 4th order polynomial to generate full RI curves within this range (**Fig. S8**)[5]. Since the thinnest PMMA thickness in Han *et al.*'s experiments was 35 nm, we first fitted the data from 35 nm to 150 nm using the polynomial (black solid line). The shaded areas represent 95% confidence intervals. The RI values for thicknesses between 15 nm to 35 nm were then extrapolated using this polynomial equation (red solid line). Other RI dependence on PMMA thickness, as reported by Unni *et al.*[6] and Todorov *et al.*[7], are shown with closed triangles.

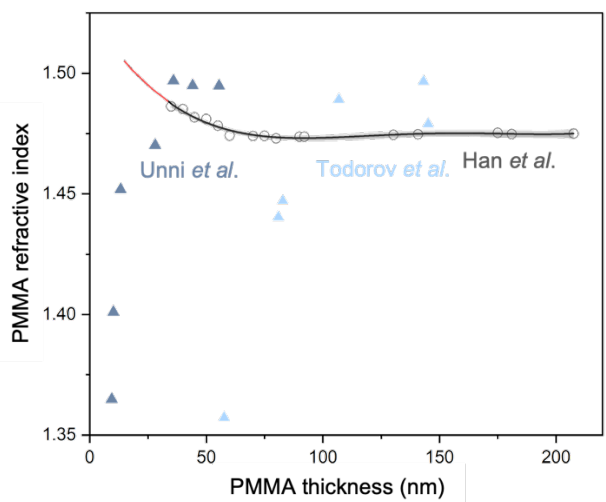

**Fig. S8** The RI data includes values from Han *et al.* (open circles, thickness > 35 nm)[5], extrapolated RI (red solid line, 15–35 nm), and data from Unni *et al.* (dark blue)[6] and Todorov *et al.* (light blue)[7].

The effective Hamaker constant for pure PMMA substance ($A_{22}$) versus polymer thickness was further calculated using Lifshitz's equation (**Eq. S2**) and the RI values in **Fig. S8**, as shown in **Fig. S9**:

$$A_{22} = \frac{3}{4}kT\left(\frac{\epsilon_2-1}{\epsilon_2+1}\right)^2 + \frac{3h\nu_e}{16\sqrt{2}}\frac{(n_2^2-1)^2}{(n_2^2+1)^{3/2}} \quad (S2)$$

where $k$ is the Boltzmann constant ($1.38 \times 10^{-23}$ J K$^{-1}$), $T$ is room temperature (298.15 K), $h$ is the Plank constant ($6.63 \times 10^{-34}$ J Hz$^{-1}$), and $\nu_e$ is the main electronic adsorption frequency constant of a material (typically $3 \times 10^{15}$ Hz)[4].

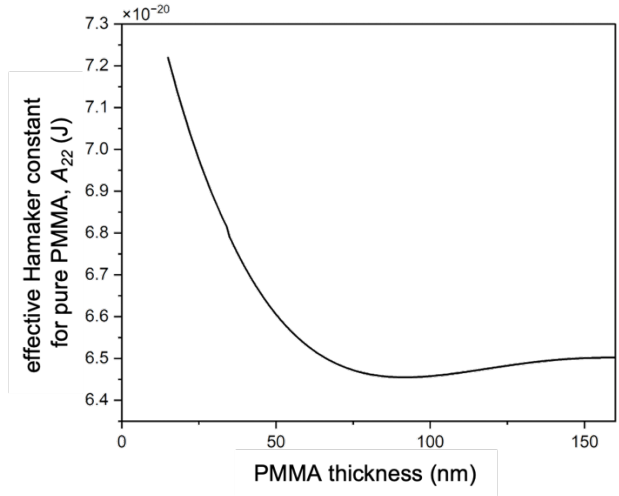

**Fig. S9** The effective Hamaker constant for pure PMMA substance ($A_{22}$) at different thickness.

The effective Hamaker constant for PMMA/Dewetting PS in air ($A_{23}$) can be further calculated by pairwise approximation, where the Hamaker constant for air ($A_{air}$) is $0^4$:

$$A_{23} = \left(\sqrt{A_{22}} - \sqrt{A_{air}}\right)\left(\sqrt{A_{33}} - \sqrt{A_{air}}\right) \tag{S3}$$

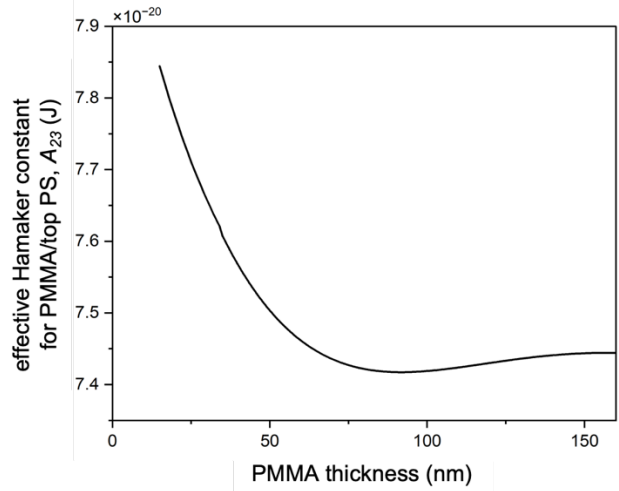

**Fig. S10** The effective Hamaker constant for PMMA/top PS ($A_{23}$) with varying PMMA thickness.